\documentclass[doublecol]{epl2}

\title{Shock-driven jamming and periodic fracture of particulate rafts}

\author{M. M. Bandi\inst{1,2} \and T. Tallinen\inst{2} \and L. Mahadevan\inst{2}}

\institute{                    
  \inst{1} CNLS and MPA-10, Los Alamos National Laboratory, Los Alamos, NM 87545, USA.\\
  \inst{2} School of Engineering and Applied Sciences, Harvard University, Cambridge, MA 02135, USA.\\
}

\abstract{A tenuous monolayer of hydrophobic particles at the air-water interface often forms a scum or raft. When such a monolayer is disturbed by the localized introduction of a surfactant droplet, a radially divergent surfactant shock front emanates from the surfactant origin and packs the particles into a jammed, compact, annular band with a packing fraction that saturates at a peak packing fraction $\phi^*$.  As the resulting two-dimensional, disordered elastic band grows with time and is driven radially outwards  by the surfactant,  it fractures to form periodic triangular cracks with robust geometrical features. We find the number of cracks $N$ and the compaction band radius $R^*$ at fracture onset vary monotonically with the initial packing fraction ($\phi_{init}$). However, its width $W^*$ is constant for all $\phi_{init}$.  A simple geometric theory that treats the compaction band as an elastic annulus, and accounts for mass conservation allows us to deduce that $N \simeq 2\pi R^*/W^* \simeq 4\pi \phi_{RCP}/\phi_{init}$, a result we verify both experimentally and numerically. We show the essential ingredients for this phenomenon are an initially low enough particulate packing fraction that allows surfactant driven advection to cause passive jamming and eventual fracture of the hydrophobic particulate interface.}

\begin{document}

\maketitle

\begin{figure}
\onefigure[width = 3.0 in]{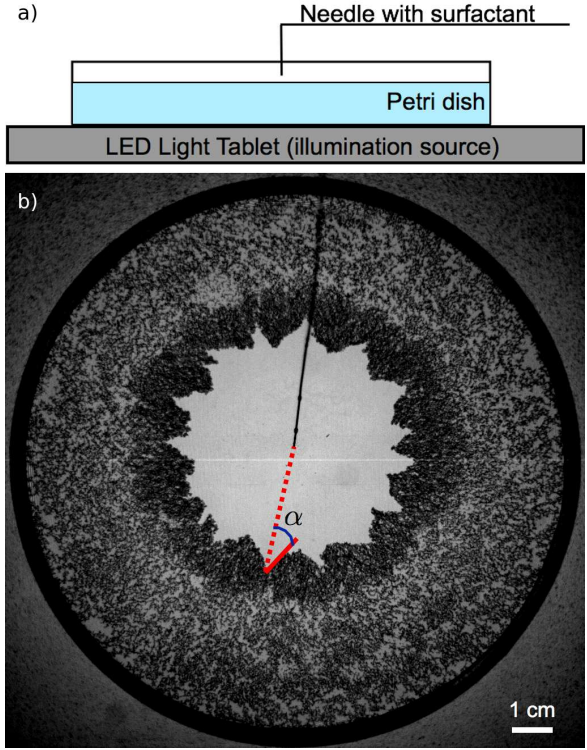}
\caption{(color online) a) Setup: A clean petri dish with distilled water is placed on a light tablet. A camera suspended vertically above records compaction dynamics of particle rafts at the air-water interface. A steel needle introduces surfactant at the interface and leads to a radially propagating surfactant shock. b) Snapshot of particle covered surface after surfactant introduction shows an annular compaction band around nearly regular triangular cracks. The compaction band's leading edge is termed the compaction front, while the surfactant shock is slightly ahead of the trailing edge. The angle $\alpha$ made by the crack face with the radial line from surfactant tip (red lines) varies within a narrow range across all cracks for all $\phi_{init}$.}
\label{fig1}
\end{figure}

The behavior of hydrophobic particles at interfaces presents interesting phenomena of fundamental significance at the intersection of interfacial physics, chemistry, and continuum mechanics \cite{Pickering, Daniels}, as well as being relevant in a variety of applications as particles at interfaces can stabilize drops and emulsions via jamming \cite{Tsapis}. Recent experiments \cite{Vella2004, Vella2009} have demonstrated the two-dimensional elastic properties of jammed hydrophobic particulate monolayers at interfaces and aspects of their deformation via buckling and random cracking. These particulate rafts also afford a nice system to study dynamics of the formation of a jammed solid, and its failure via fracture and flow, both phenomena that are hard to analyze in bulk granular media. Here we study both these phenomena at an interface using a combination of experiments and computations and show how a tenuous particulate monolayer driven by Marangoni stresses induced by localized surfactant introduction \cite{S1} leads to formation of a jammed solid and its eventual failure via a regular cracking pattern. 

Fig. \ref{fig1}a shows a schematic of the experimental setup. A clean glass petri dish (diameter 0.14 m) filled with distilled water to a height of 0.01 m is placed atop a light tablet. Teflon coated {hollow} glass particles (diameter $d = 50 \pm 10 \mu$m, specific gravity 0.25) are introduced at the air-water interface to form a particulate monolayer with an initial areal packing fraction $\phi_{init}$ (defined as ratio of total initial particulate area to total interfacial area) that varies in the range $0.1\pm 0.01 \le \phi_{init} \le 0.64 \pm 0.01$. Owing to the protocol followed for particle introduction, $\phi_{init}$ cannot be controlled, but can be measured (please see Methods section). When a clean steel needle wetted with oleic acid is dipped into the water surface at the dish center at time $t$ = 0 s, the spreading surfactant pushes the hydrophobic particles radially outwards and packs them along an annulus around the growing particle-free hole as shown in Fig. \ref{fig1}b and \cite{S1}. The compaction dynamics are imaged with a high speed digital camera (Phantom v5) at 600 frames per second {and analyzed to measure the packing fraction (see Methods section). This allows us to follow the compaction band's evolution in terms of the azimuthally averaged radial packing fraction $\phi_{\theta}(r,t) = \frac{1}{2\pi}\int_{0}^{2\pi} \phi(r,\theta,t)~d\theta$. 

Immediately following surfactant introduction, the particles move slower relative to the surfactant front due to subphase drag. After a short time, the particle dynamics settles into a self-similar form that persists for a while. Fig. \ref{fig2}a shows $\phi_{\theta}(r,t)$, normalized by the peak saturation packing fraction $\phi^*$ at various times for a representative experiment. The compaction band moves radially outwards from the location where the surfactant is introduced ($r = 0$) as time progresses \cite{S2}, with the surfactant shock front slightly ahead of the trailing edge of the band $R_T$ (see Fig. \ref{fig2} a). As the  particles ahead are swept up to form the compaction band, the peak packing fraction rises through a short transient and saturates at $\phi_{\theta}(r,t)/\phi^* \sim 1.0$  (see Fig. \ref{fig2}a). Thus, the initially tenuous low density interfacial raft gets packed by the surfactant shock and forms a jammed disordered solid when the packing fraction saturates at $\phi^*$ (at a time $t^*$) whose value was experimentally determined to be slightly below Random Close Packed Density ($\phi_{RCP} = 0.84$ in 2D \cite{OHern}). In Fig. \ref{fig2}a, we show the positions of the compaction band's leading edge $R_L$, trailing edge $R_T$, and its width $W$ as a function of time, all three of which grow with time as $t^{3/4}$ (Fig. \ref{fig2}b), a scaling that persists over almost two decades in time.

In Fig. \ref{fig2}b, where one observes $R_L$, $R_T$, and $W$ do not achieve asymptotic scaling until $t \sim 0.4t^*$. This is also supported by $\phi_{\theta}(r,t)/\phi^*$ evolution \cite{S2} which does not start rising towards peak saturation packing fraction $\phi^*$ until $t \simeq 0.4t^*$. The particles become non-inertial following this short transient after which they are passively advected by the surfactant front. However, this subtle initial effect has no bearing upon the primary experimental quantities of interest which are either static ($\phi_{init}$), or if dynamical, only become relevant at $t = t^*$ as discussed below.

The $t^{3/4}$ scaling is consistent with the classical result \cite{FayJensen} that surfactants from a constant source spread in a self-similar form in deep fluid layers i.e. the thickness of the viscous boundary layer in the fluid bulk is much smaller than the depth of the fluid layer over the duration of the experiment. This requires the ratio $\tau/T \ll 1$, where $\tau \sim 0.5$ s is the total experimental duration and $T = H^2/\nu = 100$ s ($H = 10^{-2}$ m is fluid layer depth and $\nu = 1 \times 10^{-6}$ m$^2$/s is kinematic viscosity of water) is the time required for the Blasius boundary layer to span the entire depth of the bulk fluid. The ratio $\tau/T \sim 5 \times 10^{-3}$ for our experiments, placing them in the deep fluid layer regime. The position $R_s$ of such a surfactant front for uncontaminated surfaces (please see Methods section for further details) follows the relation 
\begin{equation}
\label{eq1}
R_s = K(\frac{\Delta \gamma^2}{\mu\rho})^{1/4}t^{3/4}
\end{equation}
where $\Delta \gamma = \gamma_{(water-air)} - \gamma_{(oleic-water)} - \gamma_{(oleic-air)}$ = 23.58 $\times 10^{-3}$ N/m is the Harkins spreading coefficient at the line of three-phase contact, $\mu = 10^{-3}$ Pa$\cdot$s is the dynamic viscosity of the underlying fluid, $\rho = 10^{3}$ Kg/m$^{3}$ its density, and $K$ a numerical coefficient in the range $0.665 \le K \le 1.52$ \cite{BergTroian}. On superposing this predicted scaling for Eq. \ref{eq1} with $K=0.94$ (solid black line in Fig. \ref{fig2}b) on the experimental scaling for $R_L$ we see the particulate band propagates like the surfactant shock (Thoreau-Reynolds ridge \cite{ThoreauReynolds}), and implies the particles behave as non-inertial tracers advected by the surfactant flow consistent with earlier experiments \cite{Cazabat}. The robust $t^{3/4}$ scaling observed even after $\phi_{\theta}(r,t)/\phi^*$ saturates shows the assumption of non-inertial particle dynamics is a good one.

As particles are swept up into the compaction band  they form {a jammed disordered solid when $\phi_{\theta}(r,t)/\phi^* \rightarrow 1$ saturates at a critical time $t^*$. As the compacted band moves radially outward, it fractures to form a periodic saw-tooth pattern. There is no observable time difference between jamming and fracture onset, which is consistent with the fact that the elastic strain before fracture in the jammed solid of nearly rigid grains is likely to be very small. Fracture thus starts at $t = t^*$, and we therefore define $R^* = R_T(t^*)$ and $W^* = W(t^*)$ as the inner radius and width of the compaction annulus respectively at fracture onset.

\begin{figure}
\onefigure[width = 3.1 in]{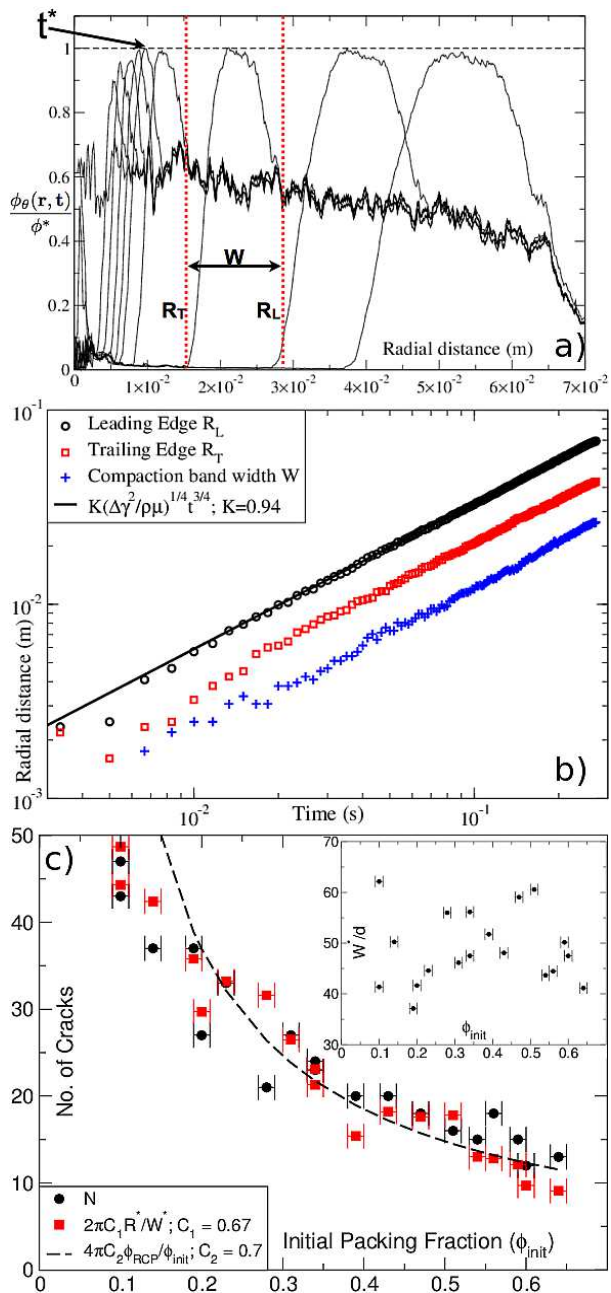}
\caption{(color online) Experiment: a) Azimuthally averaged radial packing fraction $\phi_{\theta}(r,t)/\phi^*$ vs. $r$ shows compaction shock evolution at $t$ = 0, 2, 2.3, 2.67, 3, 3.3, 4.17, 8.3, 16.7, and 25 $\times 10^{-2}$ s from left to right. Vertical dashed lines (red) mark the compaction band's shock front ($R_L$), rear ($R_T$), and width ($W$) at $t = 0.083$ s. $t^*$ (indicated at top left) represents the instant when peak packing fraction saturates at $\phi^*$. b) Position of $R_L$ (black circle), trailing edge $R_T$ (red square) and width $W$ (blue plus) vs. time t in log-log scale. The solid (black) line for $R_s$ scaling (Eq. \ref{eq1} for $K = 0.94$) exactly superimposes $R_L$ scaling. c) Measured no. of cracks $N$ (solid black circles), Eq. \ref{cracks1} (solid red squares), and Eq. \ref{cracks2} vs. $\phi_{init}$. Inset: $W^*/d$ is constant for all $\phi_{init}$.
}
\label{fig2}
\end{figure}

Beyond the time $t > t^*$, we almost never see the formation of any new cracks so that the number of cracks $N$ formed remains constant in a given experiment, even though the cracks grow dynamically. We find with increasing initial particulate packing fraction $\phi_{init}$, the number of cracks decreases monotonically (see Fig. \ref{fig2}c) as does the critical radius $R^*$. This is because the compaction annulus forms and jams at an earlier time and smaller radius with increasing $\phi_{init}$.  However the compaction band width $W^*$ exhibits no dependence on $\phi_{init}$, but as we explain later, it does depend on the particle diameter (see Fig. \ref{fig3}). Since cracks in the compaction band relieve strains over a scale comparable to the band width $W^*$, we expect that the number of cracks
\begin{equation}
N \simeq 2\pi R^*/W^*
\label{cracks1}
\end{equation}
Additionally, assuming the initial particle distribution is uniform and the jammed solid is random close packed, mass conversation dictates that the particulate area within the compaction annulus at $ t^*$ equals the particulate area within a circular radius $(R^*+W^*)$ at $t$ = 0, so ($\phi_{RCP} \pi[(R^*+W^*)^2 - (R^*)^2] = \phi_{init} \pi(R^*+W^*)^2$) so that $\phi_{RCP}/\phi_{init} = R^*/2W^*$,  thus yielding
\begin{equation}
N \simeq 2 \pi R^*/W^* \simeq 4\pi \phi_{RCP}/\phi_{init}
\label{cracks2}
\end{equation}
In Fig. \ref{fig2}c, we plot the experimentally measured values for $N$ vs.  $4\pi  C_2\phi_{RCP}/\phi_{init}$, with $C_2 = 0.7$. 

The agreement of Eq. \ref{cracks1} with experiments suggests that the continuum description holds well for this granular system, a fact also supported by Fig. \ref{fig2}c inset where $W^*/d \sim 50 \pm 10$ at all $\phi_{init}$. However, Eq. \ref{cracks2} derived from mass conservation arguments shows only partial agreement. We emphasize that this relation is based on several idealized assumptions, and holds only for intermediate values of $\phi_{init}$. Firstly, we drop the quadratic term during expansion of the mass conservation relation leading to Eq. \ref{cracks2} under the assumption that $W^* \ll R^*$ (or alternatively $N \gg 1$), which holds only when $\phi_{init} < \phi_{RCP}$. Secondly, we assume an initially uniform particle distribution. In reality, the meniscus formed by water with the petri dish wall repels hydrophobic particles towards the center. This is clearly observed as a drift in $\phi_{\theta}(r,t)$ for $r > R_L$ in Fig. \ref{fig2}a. Finally, we assume the compaction annulus jams at $\phi_{RCP}$. In reality, frictional or attractive inter-particle interactions can stabilize a granular pack below $\phi_{RCP}$. Whereas the teflon coat on particles ensures minimal inter-particulate friction, we do observe particle clustering suggesting attractive interactions are at play.

We expect both Eq. \ref{cracks1} and \ref{cracks2} will fail in the limiting cases as $\phi_{init} \rightarrow 0, \phi_{RCP}$. In the dilute limit, Eq. \ref{cracks1} and \ref{cracks2} suggest $N \rightarrow \infty$, which cannot be true.  Instead, $N$ is strongly influenced by disorder in the packing, since the crack size is only a few grain diameters, and furthermore $N \le 2\pi R^*/d$ because a crack cannot be smaller than particle diameter $d$. The continuum description however fails well before this limit is reached. In the opposite limit as $\phi_{init} \rightarrow \phi_{RCP}$, the particulate layer is either already jammed, or does not have to be packed much before it jams into a solid, so that $t^*, R^*,$ and $W^*$ are all poorly defined. Also, since particles are already constrained and no free space is available for cracks to open up and expose surfactant to the air, cracks must proceed through local re-arrangement of particles which leads to branching and kinking instabilities \cite{Vella2004}.

\begin{figure}
\begin{center}
\includegraphics[width = 3.25 in]{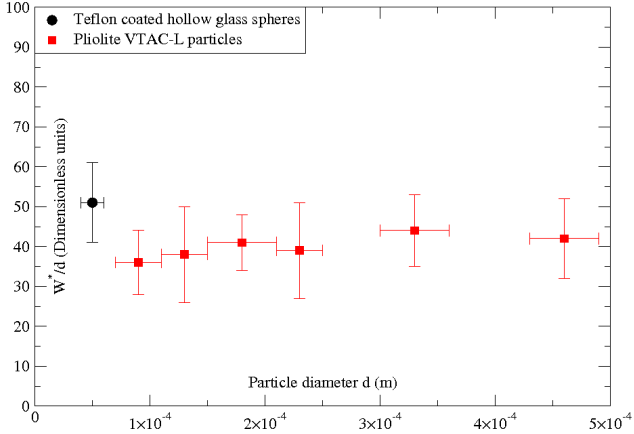}
\end{center}
\caption{(color online) Dimensionless compaction width $W^*/d$ vs. particle diameter $d$ for $d = 50 \pm 10 \mu$m (Teflon coated hollow glass spheres), $90 \pm 20 \mu$m, $130 \pm 20 \mu$ m, $180 \pm 30 \mu$m, $230 \pm 20 \mu$m, $330 \pm 30 \mu$m, and $460 \pm 30 \mu$m (all for Pliolite) shows $W^*$ scales with particle diameter. Horizontal error bars represent particle dispersity, whereas vertical error bars represent variability in measured $W^*$ over 20 experimental runs for Teflon coated hollow-glass spheres, and 4 experimental runs each for Pliolite particles.}
\label{fig3}
\end{figure}

As shown in the inset of fig. \ref{fig2}c, we found the dimensionless compaction width $W^*/d$ is independent of the initial packing fraction $\phi_{init}$. This critical width $W^*$ depends upon the particle diameter $d$, and the surface tension contrast ratio $\Delta \gamma/\gamma$ (dimensionless Marangoni stress). Here $\gamma$ is the surface tension of the oleic-air interface ($\gamma = 32.8 \times 10^{-3}$ N/m). Since the critical strain for fracture must depend on these parameters, dimensional analysis and St. Venant's principle suggest that $\epsilon = (W^*/d)(\Delta \gamma/\gamma)$. Additional experiments with varying particle diameter were performed to verify this. In addition to teflon coated hollow glass spheres, we also used polydisperse Pliolite hydrophobic particles \footnote{Pliolite VTAC-L particles from Eliokem Inc.} of various diameters. Fig. \ref{fig3} shows the dimensionless compaction width $W^*/d$ vs. particle diameter $d$ has zero slope indicating $W^*$ scales linearly with particle diameter $d$.

\begin{figure}
\onefigure[width = 3.25 in]{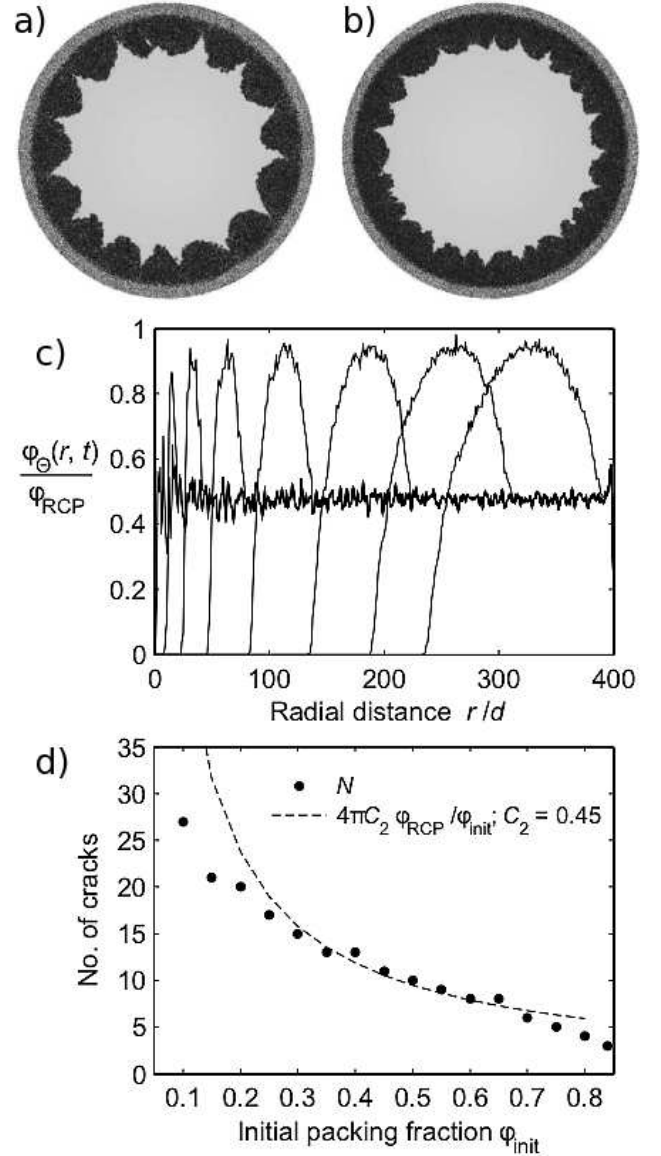}
\caption{Molecular dynamics simulation for formation and failure of a jammed solid. A compaction band is shown for (a) $\Delta\gamma = 3k/d$ and (b) $\Delta\gamma = 6k/d$. The non-dimensional system radius is $r/d = 400$ (in experiment the petri dish radius $r/d =  1400$). (c) Simulated profiles of azimuthally averaged packing fraction $\phi_{\Theta}(r, t)$ show compaction shock evolution with increasing $R_L$. (d) Simulated number of cracks is compared to Eq. \ref{cracks2} vs. $\phi_{init}$. In (c) and (d) $\Delta \gamma = 3k/d$.}
\label{fig4}
\end{figure}

Having considered the formation and number of cracks, we now analyze the crack geometry. Since all dimensions of the compaction annulus exhibit self-similar scaling (see Fig. \ref{fig2}b), the particle-free area opened by a crack too must scale self-similarly. Thus we expect the crack growth to also be self-similar, since they are driven by stresses in this growing annulus; the toothed crack front with straight edges naturally fits these constraints. An individual crack is thus all that we need characterize, and this can be done in terms of the angle ($\alpha$) between the crack face and the radial direction from the point of surfactant introduction (Fig. \ref{fig1}b) for all cracks and all values of $\phi_{init}$. The distribution of $\alpha$ is quite sharp with a coefficient of variation of $34.0^{\circ}/6.97^{\circ} = 4.9$, that is relatively constant in time as cracks grow with the diverging compaction band. 

We now turn to a quantitative description of the dynamics of compaction, jamming and fracture to complement the qualitative description of the phenomenon at a scaling level. We model the initially tenuous raft as a planar system of hard particles with a pair-wise attractive force $F = k$ at separation $r < d/10$, and $F = 0$ for $r > d/10$ with the particle diameter assumed to be uniformly distributed in the interval $[0.8d, 1.2d]$; the actual interaction potential \cite{Nikolaides} is more complicated, but the simple representation described above captures all the qualitative trends. \footnote{We also tested harmonic interaction potential without notable difference in the results.} The initial configuration is created by first placing the particles randomly and then relaxing the system which leads to some reordering initially. A symplectic Euler scheme built into a molecular dynamics simulation \cite{rapaport}  is then used to solve for the damped Newtonian dynamics of the particles. The role of fluid drag on the particles is complicated by the presence of a boundary layer. To mimic this accurately with an implicit fluid, we assume an outward radial flow with velocity $U_r = \frac{dR_s}{dt}$ inside a radius $R_s$ given by eq. (\ref{eq1}) and zero elsewhere \cite{BergTroian}, that  simulates the spreading of the surfactant and the resulting fluid flow.  We set $R_s (\hat t) = d$ initially, where $\hat t  = [d/K (\frac{ \mu \rho}{\Delta \gamma^2})^{1/4}]^{4/3}$ following eq. (\ref{eq1}), to avoid a divergence of velocity at $t = 0$; our simulations therefore start at $t = \hat t$ (the long time dynamics are independent of $\hat t$). Motion of a particle with a velocity $\bar{v}$ relative to that of fluid is opposed by a drag $\bar{f}_{\mu} = \tilde{\mu}d\bar{v}$, where we introduce an effective viscosity:
\begin{equation}
\label{mu}
\tilde{\mu} = \sqrt{\frac{\phi_{RCP} \mu\rho}{\phi_{init} t}}d.
\end{equation}
which follows from the requirement that a radial integrated pressure difference across the compaction band must be of the order $\Delta \gamma$. It can be derived by using a similar boundary layer argument that has been used to derive eq. (\ref{eq1}) itself \cite{BergTroian}. According to the Blasius boundary layer theory, drag exerted per unit width of a flat plate is $D \approx U^{3/2}\sqrt{l \mu \rho}$ where $U$ is the flow velocity and $l$ width of the plate in the direction of the flow \cite{batchelor}. Here $l = W \approx \frac{R_s \phi_{init}}{2 \phi_{RCP}}$ and $U \sim R_s/t$ yielding $D \sim \Delta\gamma \sqrt{\frac{\phi_{init}}{\phi_{RCP}}}$. On the other hand, number of particles per unit width of the band is approximated by $\frac{W}{d^2}$. Multiplying this by $\bar{f}_{\mu}$ and assuming all the particles moving with velocity $U$ leads to total drag $D_p \sim \frac{\tilde{\mu} R_s^2 \phi_{init}}{d t \phi_{RCP}}$ exerted on the raft. By requiring $D_p = D$ one obtains eq. (\ref{mu}).

In the non-inertial regime explored here, the dynamics of the model depends on the radial pressure relative to attraction between the particles, which can be controlled by $\Delta \gamma$ and $k$. For $\Delta \gamma = 3 k/d$ the model displays formation of the compaction band and its fracture in a manner similar to the experiments (Fig. \ref{fig4}a, \cite{S3}).} In the model we can accurately measure that the compacted packing fraction saturates close to $\phi_{RCP}$. Fig. \ref{fig4}c \cite{S4} shows time evolution of the azimuthally averaged packing fraction in good agreement with experiments (Fig. \ref{fig2}a). The scaling of $N$ vs. $\phi_{init}$ (Fig. \ref{fig4}d) from simulations also compares well with experiments (Fig. \ref{fig2}c), as do the deviations expected when $\phi_{init} \rightarrow 0$ and $\phi_{init} \rightarrow \phi_{RCP}$ as discussed earlier. In addition, the $N$ depends on $\Delta \gamma$ through the pre-factor $C_2$ (see Fig. \ref{fig2}c and \ref{fig4}d). The model suggests that the crack depth is controlled by the radial pressure such that for increasing $\Delta \gamma$ the crack depth decreases, see Figs. \ref{fig4}a,b. Our simulations also allow us to measure the crack angle (Fig. \ref{fig4}) $\alpha = 30.5^{\circ} \pm 3.5^{\circ}$ for $\Delta \gamma = 3k/d$ and $\alpha = 33.9^{\circ} \pm 7.2^{\circ}$ for $\Delta \gamma = 6k/d$, results which compare well with experiments.

The simple molecular dynamics scheme adopted here shows the basic features of the experimental phenomenon can be captured with a simple attractive interaction between particles, irrespective of its actual form. However, knowing the actual form of the interaction will better allow one to understand the individual and collective particle dynamics. Additionally, understanding the role of rough contact lines and particle anisotropy also requires careful future study.

Our experiments and simulations have allowed us to understand the formation and failure of the resulting compact, disordered solid in terms of a structural parameter, the initial packing fraction $\phi_{init}$, driven by a differential surface tension $\Delta \gamma$. Since the two features are common to many problems involving the mechanics of disordered materials, this system might serve as a paradigm for further studies in amorphous solids.

\section{Methods}
{\it Preparation:} Standard cleaning procedures \cite{Cleaning} were followed to ensure an impurity free setup. The petri dish was washed in dilute Sulphuric acid, rinsed in distilled water, then baked dry at 100$^{\circ}$C for 30 minutes, followed by exposure to Ultraviolet radiation in an oxygenated environment to break up residual organic impurities. The needle used for surfactant introduction was washed in Ethanol, rinsed in distilled water and flame treated prior to each experiment. All experiments reported here were performed with pure Oleic acid (no dilution with an organic solvent). The amount of surfactant introduced had no bearing upon the results.

Hollow glass microspheres composed of borosilicate glass \footnote{Trelleborg Emerson \& Cumming Eccospheres. Product No. W-25, mean size 50$\mu$m, density 0.25 g/cc, flotation 95\% bulk vol.} were coated with a thin layer of Polytetrafluoroethylene (Teflon) via molecular vapor deposition. Prior to the experiment, the particles were washed in Ethanol and rinsed with distilled water and baked dry at 100$^{\circ}$C. Particles were introduced by puffing them in air and allowing them to naturally settle onto the interface. The initial packing fraction $\phi_{init}$ cannot be controlled in this particle deposition scheme. A number of experimental runs were performed by varying the approximate number of particles released which indirectly permitted us to span a range of $\phi_{init}$.

{\it Surfactant:} The validity of Eq. \ref{eq1} requires that the propagating surfactant front be introduced from a source of constant concentration. This requirement was experimentally met by introducing the surfactant from a point source (needle) with a concentration well in excess of the surfactant's Critical Micelle Concentration (CMC), which also ensured the surfactant surface tension $\gamma$ remained constant across all experimental runs and was not a function of surfactant concentration as also confirmed in \cite{Vella2004}. For an Oleic acid molecule with cross-sectional area of 20 \AA$^2$, the total number of molecules required to form a mono-molecular layer at the petri dish's air-water interface is $7.7 \times 10^{16}$ molecules which translates to an approximate total Oleic acid mass of 0.36 $\mu$g (at Oleic acid density of 282.46 g/mole). The average droplet mass introduced by the needle was 7.66 mg ($\sim 8.5~\mu$l volume per droplet), more than $2 \times 10^{4}$ times the quantity required to form a monolayer. Hence we are confident the surfactant spreads as a thick surfactant layer and its surface tension is concentration independent. We also conducted five independent measurements each for the surface tension at the air-water, air-oleic acid, and water-oleic acid film (formed by introducing droplet with the needle) interfaces using the Wilhelmy plate method which confirmed the surface tension for surfactant film was concentration independent.

{\it Image Analysis:} The high speed camera (Phantom v5.0 camera, exposure time: 150 $\mu$s with 28 mm Nikkor manual focus wide angle lens, aperture setting: f/5.6) recorded light transmitted through the particulate layer providing high contrast images of dark particles in a bright background. All images were collected under same illumination conditions i.e. DC illumination source intensity, camera lens aperture, and exposure time were kept constant across all runs. A digital snapshot of the background (petri dish with distilled water prior to particle introduction) was subtracted from images with particle dynamics thereby removing inevitable minor spatial illumination inhomogeneities. Image analysis algorithms for particulate area measurement were developed in-house and first calibrated against particles of known area (measured under a microscope) that were introduced at the interface. Tests performed against images obtained from molecular dynamics simulations provided the error bars we present in Fig. \ref{fig2}. The measurement error for $\phi_{init}$ was estimated at $\delta \phi_{init} = 0.01$. For dynamical measurements, control tests against molecular dynamics images yielded a linear increase in error for $\phi > 0.7$ reaching a peak error $\delta \phi_{\theta}(r,t) = 0.04$. Given this higher error at high packing fractions, we are unable to confirm whether the compaction band saturates at $\phi_{RCP} = 0.84$ or at a lower value due to attractive interactions as discussed earlier. In any event, this has no bearing upon the results since we only seek to learn the instant $t^*$ when the peak packing fraction saturates to a maximum value $\phi^*$ heralding the formation of a jammed solid.

\acknowledgments
This work was supported by the U.S. Dept. of Energy at Los Alamos National Laboratory under Contract No. DE-AC52-06NA25396 (MMB), the Harvard NSF-MRSEC and the MacArthur Foundation (LM). The authors acknowledge WI Goldburg, MK Rivera, and RE Ecke for equipment support during preliminary investigations, and D Vella and A Shreve for discussions.

\end{document}